\documentclass[pra,superscriptaddress,twocolumn,showpacs]{revtex4-1}
\usepackage{graphicx}% Include figure files
\usepackage{dcolumn}% Align table columns on decimal point
\usepackage{bm}% bold math
\usepackage{color}% couleur en rgb

\begin{document}
\title{Photon Self-Induced Spin to Orbital Conversion in TGG crystal at high laser power.}
\author{S. Mosca}
\affiliation{Dipartimento di Scienze Fisiche, Universit\`{a} di
Napoli ``Federico II'', Compl.\ Univ.\ di Monte S. Angelo, 80126
Napoli, Italy}
\affiliation{INFN Sezione di Napoli, Compl.\ Univ.\ di Monte S. Angelo, 80126 Napoli, Italy}
\author{B. Canuel}
\affiliation{EGO, European Gravitational Observatory,
Via E. Amaldi, 56021 S. Stefano a Macerata - Cascina (PI)
ITALY}
\author{E. Karimi}
\affiliation{Dipartimento di Scienze Fisiche, Universit\`{a} di
Napoli ``Federico II'', Compl.\ Univ.\ di Monte S. Angelo, 80126
Napoli, Italy}
\author{B. Piccirillo}
\affiliation{Dipartimento di Scienze Fisiche, Universit\`{a} di
Napoli ``Federico II'', Compl.\ Univ.\ di Monte S. Angelo, 80126
Napoli, Italy}
\affiliation{CNISM-Consorzio Nazionale Interuniversitario per le Scienze
Fisiche della Materia, Napoli}
\author{L. Marrucci}
\affiliation{Dipartimento di Scienze Fisiche, Universit\`{a} di
Napoli ``Federico II'', Compl.\ Univ.\ di Monte S. Angelo, 80126
Napoli, Italy}
\address{CNR-INFM Coherentia, Compl.\ Univ.\ di Monte S.
Angelo, 80126 Napoli, Italy}
\author{R. De Rosa}
\affiliation{Dipartimento di Scienze Fisiche, Universit\`{a} di
Napoli ``Federico II'', Compl.\ Univ.\ di Monte S. Angelo, 80126
Napoli, Italy}
\affiliation{INFN Sezione di Napoli, Compl.\ Univ.\ di Monte S. Angelo, 80126 Napoli, Italy}
\author{E. Genin}
\affiliation{EGO, European Gravitational Observatory,
Via E. Amaldi, 56021 S. Stefano a Macerata - Cascina (PI)
ITALY}
\author{L. Milano}
\affiliation{Dipartimento di Scienze Fisiche, Universit\`{a} di
Napoli ``Federico II'', Compl.\ Univ.\ di Monte S. Angelo, 80126
Napoli, Italy}
\affiliation{INFN Sezione di Napoli, Compl.\ Univ.\ di Monte S. Angelo, 80126 Napoli, Italy}
\author{E. Santamato}\email{enrico.santamato@na.infn.it}
\affiliation{Dipartimento di Scienze Fisiche, Universit\`{a} di
Napoli ``Federico II'', Compl.\ Univ.\ di Monte S. Angelo, 80126
Napoli, Italy}
\affiliation{CNISM-Consorzio Nazionale Interuniversitario per le Scienze
Fisiche della Materia, Napoli}
%%%
\date{July 2010}
%%%
\begin{abstract}
In this paper, we present experimental evidence of a newly discovered third-order nonlinear optical process Self-Induced Spin-to-Orbital Conversion (SISTOC) of the photon angular momentum. This effect is the physical mechanism at the origin of the depolarization of very intense laser beams propagating in isotropic materials. The SISTOC process, like self-focusing, is triggered by laser heating leading to a radial temperature gradient in the medium. In this work we tested the occurrence of SISTOC in a terbium gallium garnet (TGG) rod for an impinging laser power of about 100~W. To study the SISTOC process we used different techniques: polarization analysis, interferometry and tomography of the photon orbital angular momentum. Our results confirm, in particular, that the apparent depolarization of the beam is due to the occurrence of maximal entanglement between the spin and orbital angular momentum of the photons undergoing the SISTOC process. This explanation of the true nature of the depolarization mechanism could be of some help in finding novel methods to reduce or to compensate for this usually unwanted depolarization effect in all cases where very high laser power and good beam quality are required.
\end{abstract}
\maketitle
There are several optics experiments in the world where both high laser power and excellent beam quality are simultaneously mandatory. For instance, this is the case for the next generation of optical interferometers used to detect gravitational waves, where high laser power, of the order of 200 W, in the fundamental mode, is required to increase the detector sensitivity \cite{virgo1,ligo1,GEOHF}. Thermal effects appearing in high power lasers or in the bulk optical components exposed to high laser power due to non-negligible absorption can strongly affect the beam quality. In solid state lasers, thermal gradients within the laser medium cause refractive index changes leading to thermal lensing, aberrations and birefringence. In particular, the thermally induced birefringence is known to introduce a depolarization of the light that becomes the limiting effect on power scaling \cite{fortster70,mourdough96,frede04}. Ways to compensate for this effect have been recently proposed~\cite{Khazanov99, frede04, frede04a}, thus opening the possibility to realize high power and high quality continuous wave lasers that could be used in gravitational wave interferometers or in other applications. The Faraday Isolator (FI) is one of the components to be most strongly affected by thermal effects and is fundamental for the success of gravitational wave optical experiments. The magneto-optic crystal used in the FI is Terbium Gallium Garnet (TGG) which has a relatively high absorption (generally higher than 1000~ppm$\cdot$cm$^{-1}$)~\cite{zarubina92}. This absorption creates an overall temperature increase of the magneto-optic crystal and generates a non-uniform temperature distribution over the transverse cross section of the optical element.  Both effects can significantly impact Faraday isolation when going to high power. The first effect is associated with the Verdet constant change and is detailed in~\cite{AO1}. The second effect is associated to mechanical stresses induced by temperature gradient and gives the main contribution to the apparent depolarization of high power laser beams and to the consequent deterioration of the degree of isolation~\cite{Khazanov99}.\\

In the present paper, we aim to demonstrate that the mechanism creating the depolarization and therefore spoiling the Faraday isolator performance is a self-induced partial Spin-To-Orbital Conversion (STOC) of the input photon Spin Angular Momentum (SAM) into the Orbital Angular Momentum (OAM)~\cite{marrucci06, karimi09,karimi09a}. The Self-Induced STOC, or SISTOC, in fact, may put some photons of the input beam into particular states where the photon SAM and OAM are maximally entangled. The apparent depolarization of the beam is a direct manifestation of the decoherence of two quantum degrees of freedom (the photon SAM, here) when they are entangled. The depolarization is said to be ``apparent'', here, because it could be removed, for example, by a quantum erasing apparatus~\cite{erasing}.

In order to gain a deeper insight into the SISTOC process we used three different techniques: polarimetry, interferometry and full OAM tomography in the photon spinorbit space. Our experiments were carried out with the input laser beam polarized either linearly or circularly. As expected from the theory, SAM-OAM entanglement and apparent depolarization of the photons converted by SISTOC was only found in the case of linearly polarized incident beam. In the case of the circular polarization, there is no SAM-OAM entanglement and the SISTOC converted photons were found to be in OAM eigenstates with full circularly polarization with helicity depending on the helicity of the input beam.
\subsection{Photo-elastic effect by radial thermal gradient and induced birefringence}
\begin{figure}[htb]
	\includegraphics[width=6cm]{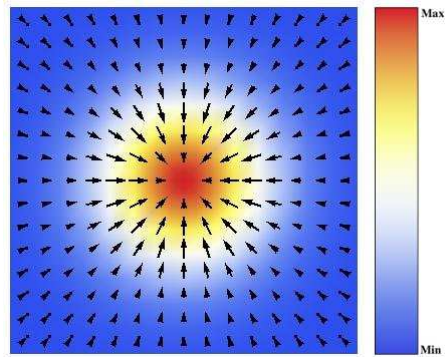}
  	\caption{\label{fig:heatedTGG} Temperature distribution in the TGG transverse plane. The arrows show the temperature gradient as well as the local direction of the thermally induced birefringence axis.}
\end{figure}

In the case of a high intensity TEM$_{00}$ Gaussian beam, the optical material experiences temperature gradients imprinted by the bell-shaped beam profile. These gradients introduce a mechanical stress in the material creating a birefringence with a radial symmetry. In isotropic materials~\cite{anisotropy}, the birefringence axis follows the radial direction, along the temperature gradient and the birefringence optical retardation $\delta(r)$ depends only on the radial coordinate $r$. The temperature-induced birefringence retardation $\delta(r)$ can be found by solving the thermal and elastic problem in the material~\cite{Khazanov99}, \cite{khazanov04}. In the case of a Gaussian heat source, $\delta(r)$ is given by
\begin{equation}\label{eq:delta}
  	\delta(r)=\frac{\alpha P_0 L \Upsilon }{\kappa \lambda}\left(1+\frac{-1+e^{-2r^2/r_0^2}}{2r^2/r_0^2}\right),
\end{equation}
where $\lambda$ is the optical wavelength, $r_0$ is the beam radius at the rod position, $\alpha$ and $\kappa$ are respectively the absorption coefficient and the thermal conductivity of the material, and $\Upsilon$ is an effective opto-elastic coefficient. Taking typical values for TGG~\cite{khazanov04} $\Upsilon=4,7\cdot10^{-6}$~K$^{-1}$, $\alpha=1500$~ppm~cm$^{-1}$, $\kappa=7.4$~W~m$^{-1}$~K$^{-1}$, $L=18$~mm, and considering an incident power $P_0=125$~W, we found $\frac{\delta}{2}\simeq 5.7\ensuremath{^\circ}$. We may then consider $\frac{\delta(r)}{2}$ in Eq.~(\ref{eq:eout_circ}) below as a small quantity. However, the key point in producing the SISTOC effect is the radial direction of the local optical axis in the heated material, as shown in Fig.~\ref{fig:heatedTGG}. In fact, the indefinite birefringence direction located at the center of the heated medium creates a topological singularity of charge $q=1$ which is tranferred into the phase of the optical beam leading to a vortex light beam carrying OAM. We may regard the heated material as a $q$-plate, an optical device recently developed for orbital angular momentum (OAM) manipulation exploiting the STOC process~\cite{marrucci06}.
The local orientation $\psi$ of the optical axis in the plane of the $q$-plate is generally written as  $\psi=q\phi+\alpha_0$ \cite{marrucci06}, where $\phi$ is the azimuthal angle, leading to the analogy between the $q$-plate and the heated TGG for $q=1$ and $\alpha_0=0$. $Q$-plates are usually made with liquid materials and their optical retardation $\delta$ is uniform, whereas the optical retardation of the heated TGG has the radial distribution (\ref{eq:delta}). The effect of the heated TGG on a polarized beam can be calculated in the same way as for the $q$-plate, using the Jones matrix formalism and taking into account the radial dependence of the optical delay.
\subsection{SISTOC process}
The Jones matrix $\hat F(r,\phi)$ representing the heated material is given by~\cite{marrucci06}
\begin{equation}\label{eq:F}
  \hat{F}(r,\phi)= \hat{R}(\phi)\left( \begin{array}{cc}
                                              e^{-\frac{{\rm i} \delta(r)}{2}} & 0 \\
                                              0 & e^{\frac{{\rm i} \delta(r)}{2}}
                                      \end{array}
                                \right)\hat{R}(-\phi),
\end{equation}
where $\delta(r)$ is given by Eq.~(\ref{eq:delta}) and $\hat{R}$ is the rotation matrix of angle $\phi$
\begin{equation}
  	\hat{R} = \left( \begin{array}{cc}
                \cos\phi & \sin\phi \\
                -\sin\phi & \cos\phi
            \end{array} \right)
\end{equation}
\\

Let us consider a high power TEM$_{00}$ Gaussian beam impinging on the TGG rod. In the case of the circular polarization, the input optical field is ${\bm E}_{in}=E_0(r){\bm e}_\pm$, where ${\bm e}_{\pm}=\left({\bm e}_x \pm {{\rm i}}{\bm e}_y \right)/ \sqrt{2}$. The output field $\bm E_{out}(r,\phi)=\hat{F}(r,\phi){\bm E}_{in}$ transmitted beyond the heated rod is calculated from Eq.~(\ref{eq:F}) as
\begin{equation}\label{eq:eout_circ}
     	{\bm E}_{\rm out}(r,\phi)=E_0(r)\left[\cos\frac{\delta(r)}{2}{\bm e}_\pm
           - {\rm i} \sin\frac{\delta(r)}{2}e^{\pm2{\rm i}\phi}{\bm e}_\mp\right]
\end{equation}
The first term on the right of Eq.~(\ref{eq:eout_circ}) proportional to $\cos\delta/2$ has the same circular polarization and radial dependence of the input field. We will refer to this term as to the unconverted part of the input beam. The unconverted part of the beam carries no OAM. The second term, proportional to $\sin\delta/2$ has opposite circular helicity and presents the characteristic phase factor $\exp(\pm2{\rm i}\phi)$, corresponding to the definite OAM content of $\pm2\hbar$ per photon. This term is the part of the input beam that was converted by the SISTOC process. Each photon in the converted part of the beam has its SAM changed by $\mp2\hbar$ and its OAM changed by $\pm2\hbar$, thus leaving the total (SAM+OAM) angular momentum conserved, which is a peculiar feature of the STOC process~\cite{marrucci06}. From Eq.~(\ref{eq:eout_circ}) we see that $\delta(r)$ and, hence, the fraction of photons that are converted by STOC, depends on the power $P_0$ carried by the beam itself, which is the characteristic of self-induced third-order nonlinear optical process. We may then regard the SISTOC as a thermally induced nonlinear process as self-focusing, but able to change the OAM content of the beam.\\

In the case of the input beam being linearly polarized along the $x$ axis the input optical field is ${\bm E}_{in}=E_0(r){\bm e}_x$, and the field transmitted beyond the heated material is given by
\begin{eqnarray}\label{eq:eout_lin}
    	{\bm E}_{\rm out}(r,\phi)=E_0(r)\left[\cos\frac{\delta(r)}{2}\bm{e}_x\, -\right. \nonumber\\
    \left.{\rm i}\sin\frac{\delta(r)}{2}\left(\cos{2\phi}\;{\bm e}_x + \sin{2\phi}\;{\bm e}_y \right)\right].
\end{eqnarray}
From Eq.~(\ref{eq:eout_lin}) we see once again that the term proportional to $\cos\delta/2$ is the unconverted part of the input beam, while the other term, proportional to $\sin\delta/2$, is the part converted by SISTOC. However, unlike the previous case, where the SISTOC converted photons were circularly polarized, in this case the converted part is a coherent, maximally \textit{entangled}, superposition of the left- and right-circular polarizations and the $\pm2\hbar$ eigenstates of the photon OAM. It is precisely this spinorbit entanglement the ultimate reason of the apparent complete depolarization~\cite{reducedmatrix} of the converted field noticed in previous works~\cite{Khazanov99}.\\

\begin{figure}[thb]
	\includegraphics[width=8.5cm]{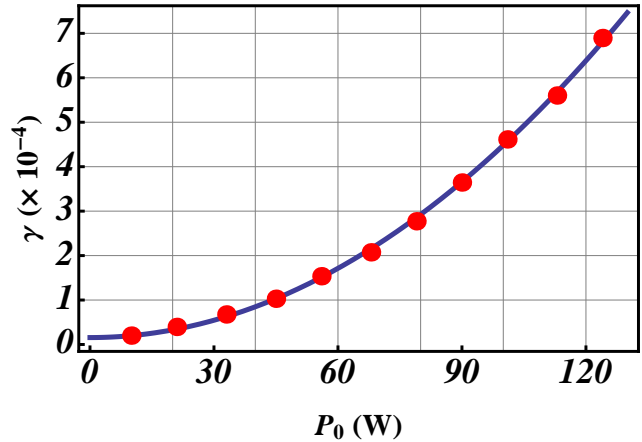}
  	\caption{\label{fig:gamma} Fraction $\gamma$ of the incident power converted by the SISTOC process as a function of the incident power $P_0$. The fit (solid curve) confirms the square dependence.}
\end{figure}

Our measurements were carried out in the far field beyond the heated TGG material. In order to compare the experimental data with theory, we must Fourier-transform the field given in Eqs.~(\ref{eq:eout_circ}) and (\ref{eq:eout_lin}). For a Gaussian input field, the result is
\begin{eqnarray}\label{eq:farfield}
 	{\bm E}_{\rm SISTOC}^{\rm far}=\frac{3\pi\alpha P_0 L \Upsilon r_0^2}{\kappa\lambda}\frac{{\rm e}^{-\frac{3 \rho^2}{2}}}{\rho^2}
          \left(1-{\rm e}^{\rho^2}+\rho^2\right) \times \nonumber \\
     \times\left\{ \begin{array}{cc}
                  e^{2{\rm i}\phi}{\bm e}_- \\
                  (\cos 2\phi\;{\bm e}_x + \sin 2\phi\; {\bm e}_y)
              \end{array}
     \right.,
\end{eqnarray}
where $\rho=(\sqrt{2}\pi w_0/\sqrt{3}\lambda z)r'$, and $r'$ is the radial coordinate in the far field transverse plane ad distance $z$. In calculating Eq.~(\ref{eq:farfield}), we considered only the SISTOC part of the output beam in the limit of small $\delta$. The upper row in Eq.~(\ref{eq:farfield}) refers to the case of left circular polarization of the input beam and the lower row the case of the linear polarization along $x$. In Fig.~\ref{fig:farfield} we show the calculated far-field transverse profiles of the SISTOC part of the beam. In the case of input left circular polarization (Fig.~\ref{fig:farfield}a), the intensity takes the typical doughnut profile of the OAM eigenstates. In the case of input linear polarization the intensity profile of the $x$ and $y$ components of the far field have four-leaf clover shapes rotated by 45\ensuremath{^\circ} each other (Fig.~\ref{fig:farfield}b,c, respectively). Finally, we notice that in view of Eq.~(\ref{eq:farfield}) the power fraction converted by the SISTOC process scales as the square of the incident power, as shown in Fig.~\ref{fig:gamma} and found in previous experiments~\cite{Khazanov99}.
\subsection{The experimental study of SISTOC in a TGG crystal}
\begin{figure}[thb]
 	\includegraphics[width=8.5cm]{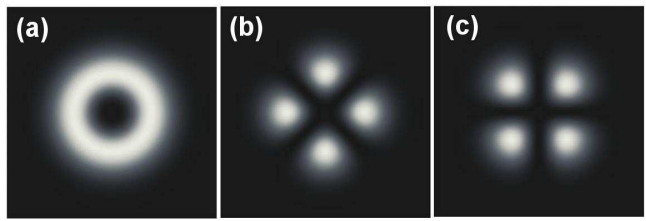}
  	\caption{\label{fig:farfield}Expected far-field beam profile of SISTOC converted modes calculated from Eq.~(\ref{eq:farfield}). Left-circular input polarization (a); $x$-linear input polarization $x$-component (b);  $y$-component (c).}
\end{figure}
\begin{figure}[htb]
	\includegraphics[width=8.5cm]{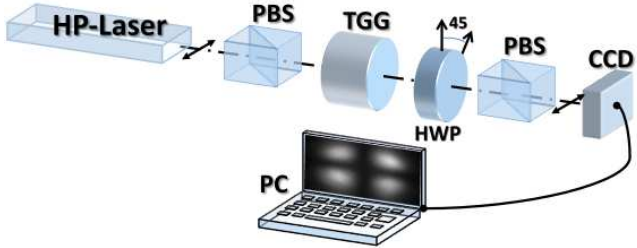}
	\caption{\label{fig:lin_depol}Measurement of the intensity mode profile of the SISTOC light in TGG with the linear input polarization at $P_0=125$~W. A half-wave plate placed in front of the second PBS is adjusted to obtain minimum of transmission beyond the second polarizer, in order to have only the SISTOC light transmitted. A four-leaf clover shape is observed beyond the second PBS.}
\end{figure}
\begin{figure}[htp]
	\centering
	\includegraphics[width=8.5cm]{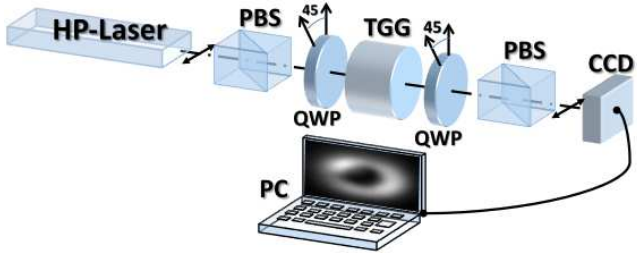}
	\caption{\label{fig:circ_depol}Measurement of the intensity mode profile of the SISTOC light in TGG with the circular input polarization at $P_0=125$~W. The first quarter-wave plate makes the polarization left circular on the TGG material and the second quarter-wave plate at $45\ensuremath{^\circ}$ turns back the polarization into linear along $y$ to be reflected by the second PBS. A doughnut-shaped vortex beam is observed beyond the second PBS.}
\end{figure}
\begin{figure}[thp]
	\centering
	\includegraphics[width=8.5cm]{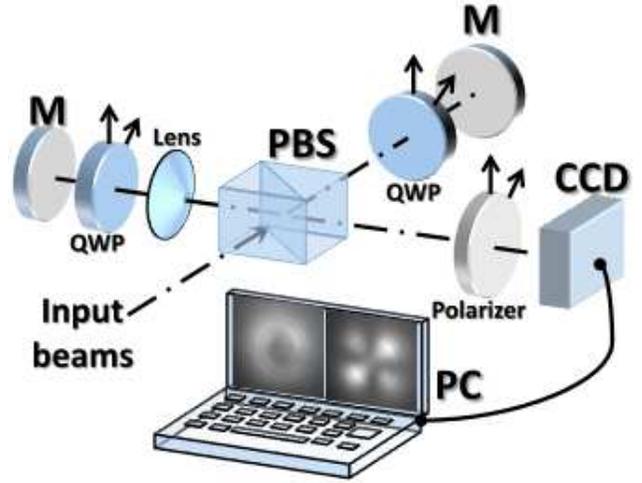}
	\caption{\label{fig:interf}The SISTOC light was isolated as in the previous experiments. The second polarizer, however, transmits a small part of the unconverted light too. The converted and unconverted fields sent into the interferometer have orthogonal polarizations. The unconverted light is used as reference beam and sent into the upper arm of the interferometer. The reference beam passes twice through a quarter-wave plate to send back the whole field to the output of the interferometer. The mirror of this arm is placed on a translation stage to adapt the differential length of the two arms. A lens makes the reference beam wavefront curved. The SISTOC beam is sent into the other arm of the interferometer. The reference and the SISTOC beam interfere in the output polarizer at 45\ensuremath{^\circ}. The fringe pattern is observed by the CCD camera.}
\end{figure}
\begin{figure}[thp]
	\centering
	\includegraphics[width=8.5cm]{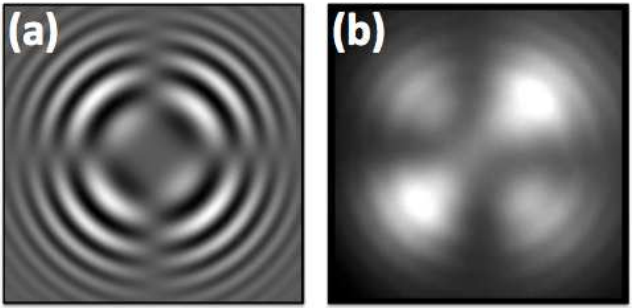}
	\caption{\label{fig:interf_lin} (a) the simulated pattern; (b) interference pattern obtained on the CCD camera of the setup shown in Fig.~\ref{fig:interf} for $P_0=150$~W and linear input polarization along $x$. The $\pi$-phase shift between each lob of the four-leaf clover may be noticed.}
\end{figure}

In our experiments we used a $<$111$>$-cut TGG rod (diameter=20~mm, length=18~mm) from Northrop Grumman with an absorption of 1900 ppm~cm$^{-1}$. The laser source was a high power diode-pumped Ytterbium fibre laser from IPG photonics delivering up to 200 W at a wavelength close to 1064 nm. We setup several experiments to analyze the SISTOC part of the beam emerging form the heated TGG material. When the input beam was linearly polarized along $x$, we isolated the converted part of the output beam using a half-wave plate at $45\ensuremath{^\circ}$ and a linear polarizer aligned along $x$~(see Fig.~\ref{fig:lin_depol}). Only the $y$-component of the emerging field was analyzed, because the $x$-component was overwhelmed by the unconverted part of the input beam. The far-field intensity profiles of the radiation converted by SISTOC was detected by a CCD camera. The experimental intensity profiles are shown in Fig.~\ref{fig:lin_depol} for the case of linear input polarization and in Fig.~\ref{fig:circ_depol} for the case of circular input polarization. As we see, the observed patterns in Figs.~\ref{fig:lin_depol} and \ref{fig:circ_depol} are in very good agreement with Figs.~\ref{fig:farfield}(c) and \ref{fig:farfield}(a), respectively. The transition from the doughnut to the four-leaf clover profile is evident.
To prove that the SISTOC converted beam acquires the phase azimuthal dependence $\exp(2i\phi)$ corresponding to the OAM eigenvalue $2\hbar$ per photon, we arranged the interferometer shown in Fig.~(\ref{fig:interf}).
To isolate the SISTOC converted field, we used the same setups shown in Figs.~\ref{fig:circ_depol} and \ref{fig:lin_depol}. The only difference was the polarization quality of the second PBS, voluntarily chosen as worst so to transmit, together with the converted field generated by the heated material, a small fraction (about $\varepsilon$=1/1000 intensity) of the $x$-polarized unconverted light. Therefore, at the entry of the interferometer, we have the sum of two fields into orthogonal polarizations, viz.
\begin{equation}\label{eq:einterf}
	\bm E_{\rm interf} = \sqrt{\varepsilon} E_{0}(r) {\bm e}_x + (\bm E_{\rm SISTOC}(r,\phi)\cdot\bm e_y) {\bm e}_y.
\end{equation}
where $\bm E_{\rm SISTOC}(r,\phi)$ is given by Eq.~(\ref{eq:farfield}). It is worth noting that the two terms on the right of Eq.~(\ref{eq:einterf}) have a different OAM content and different polarization. The first term was unconverted by SISTOC and is left in the TEM$_{00}$ mode with no OAM, while the second term is converted into a linear combination of the OAM eigenvalues $\pm2\hbar$, in general. The unconverted part of the field was used as reference and it was sent into the upper arm of the interferometer and the SISTOC converted field was sent into the other arm. The two fields were made to interfere in the polarizer oriented at 45\ensuremath{^\circ} and the fringe pattern was detected by the CCD camera. The observed interference pattern for the linear input polarization is shown in Fig.~(\ref{fig:interf_lin}) on the right. The interference pattern reveals a $\pi$-phase-shift between each lob of the four-leaf clover. This pattern is completely in agreement with that calculated from Eq.~(\ref{eq:farfield}) and shown on the left of Fig.~(\ref{fig:interf_lin}).\\

\begin{figure}[htp]
	\centering
	\includegraphics[width=8.5cm]{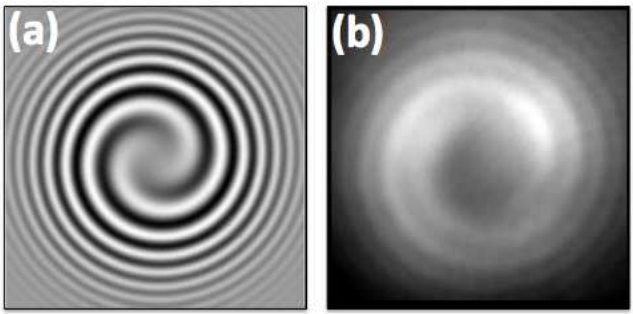}
	\caption{\label{fig:interf_circ}(a) the simulated pattern; (b) interference pattern obtained on the CCD camera of the setup shown in Fig.~\ref{fig:interf} for $P_0=100$~W and  left circular input polarization. The two-branch spiral shape characteristic of the OAM eigenvalue $2\hbar$ may be noticed.}
\end{figure}

\begin{figure}[htp]
	\centering
	\includegraphics[width=8.5cm]{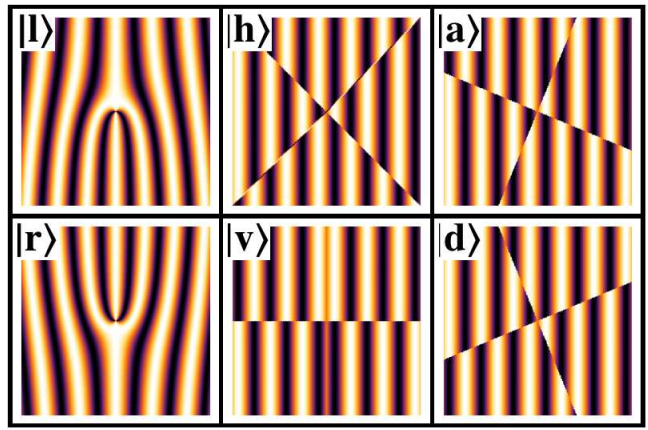}
	\caption{\label{fig:holo} Patterns of the holograms used in the experiment. The holograms in the same column correspond to the two eigenstates of Pauli's operators $\hat\sigma_z, \hat\sigma_x,\hat\sigma_y$ in the 2D Hilbert space spanned by the OAM eigenstates of $\hat\sigma_z$. Each hologram corresponds to the OAM state reported in the upper left corner. The state notation is the same as used for the photon SAM, but the corresponding symbols are put in the lower case to indicate OAM.}
\end{figure}

\begin{figure}[htp]
	\centering
	\includegraphics[width=8.5cm]{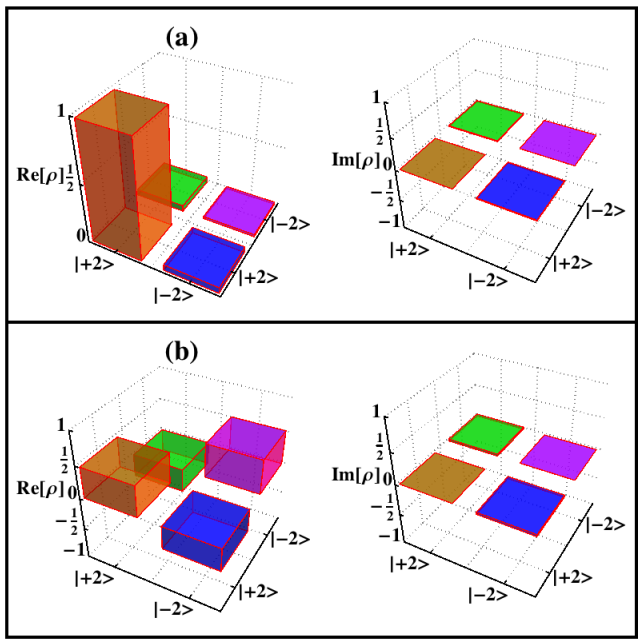}
	\caption{\label{fig:tomo} Experimental density matrix for (a) left circular input polarization and (b) for linear input 	polarization along $x$. The left and the right charts show the real and imaginary parts, respectively. The OAM eigenvalues are in units of $\hbar$.}
\end{figure}

The calculated and observed interference patterns for the circular input polarization respectively are shown on the left and right sections of Fig.~(\ref{fig:interf_circ}). In this case, the two patterns exhibit the two-branch spiral shape characteristic of the OAM eigenvalue $2\hbar$.
To complete the analysis of the OAM content of the SISTOC converted light, we made a full OAM tomography of the SISTOC light generated by the heated TGG rod. The main advantage of the OAM tomography is its capability to measure both amplitude and relative phase of the OAM components of a light beam~\cite{padgett99}. In our case, the polarization of the collected SISTOC light is fixed to be orthogonal to the input beam, so only the tomography of the OAM content of the beam is required. Moreover, the cylindrical symmetry of the system, allows us to restrict the tomography to the two-dimensional Hilbert space spanned by the eigenvectors of the photon OAM with opposite eigenvalues $\pm2\hbar$. In our experiment, we used the holographic tomography technique~\cite{padgett99, jones01} already used in single photon OAM-based experiments~\cite{nagali09, nagali09a}. This technique exploits the six computer generated holograms shown in Fig.~\ref{fig:holo}.
We made these holograms by a photographic technique, starting from computed images. After chemical bleaching, the first-order diffraction efficiency of our holograms was about 10\% at 1064~nm wavelength. The measurements were made by carefully aligning each hologram with the SISTOC beam transmitted by the heated TGG rod and collecting the ``TEM$_{00}$-like'' spot of the far field in the first-order diffraction direction~\cite{padgett99}. The far field spot was collected at the focal plane of a microscope objective and filtered through a small aperture iris. This technique, in fact, projects any unknown photon OAM state on the OAM state fixed by the hologram~\cite{padgett99}. The 2D OAM subspace considered here is isomorphic to the 2D space of the photon spin. We may think of the holograms in Fig.~\ref{fig:holo} as equivalent to polarizers acting in the spin space. The holograms in the first column correspond to polarizers selecting the left (l) and right (r) circular polarizations; the holograms in the second column correspond to polarizers selecting the horizontal (h) and vertical (v) polarizations; the holograms in the third column correspond to polarizers selecting the antidiagonal (a) and diagonal (d) polarizations, as indicated in the upper left corner of the images. In complete analogy to the polarization state analysis, we measured the ``Stokes-like'' parameters $s_3$ in each one of the three above mentioned bases so to reconstruct the density matrix of the OAM state~\cite{jones01}. The real and imaginary parts of the density matrix obtained from our measurements are shown in Fig.~\ref{fig:tomo} for the circular input polarization (a) and the linear input polarization (b).
The fidelity of SISTOC process for the case of circular and linear input polarizations are $0.98$ and $0.86$, respectively, which are promising results and show that our experimental results are in a very good agreement with our theoretical model~\cite{fidelity}. As expected from Eq.~\ref{eq:farfield}, in the case of the circular input polarization, the SISTOC converted photons are put into the OAM eigenstate $2\hbar$, while in the case of the linear input polarization, the state of the $y$-component of the SISTOC photons is an equally-weighted antisymmetric superposition of the OAM eigenstates $2\hbar$ and $-2\hbar$.
\subsection{Conclusions}
We have proved by a series of experiments, including full OAM state tomography that the apparent depolarization observed when a very high power laser beam passes through a medium is due to a new thermally induced third-order process, namely Self-Induced Spin-To-Orbital Conversion (SISTOC), where a power-dependent fraction of the incident photons converts its angular momentum from spin into orbital. Our experiments are in full agreement with a model where the SISTOC conversion is limited to the 2D OAM subspace spanned by the OAM eigenstates $\pm2\hbar$ per photon. The SISTOC process is triggered by the birefringence induced in the material by radial temperature gradient due to light absorption. The fraction of light suffering the SISTOC process remains proportional to the square of the input laser power up to about 100~W. The light depolarization is apparent because it is not due to random dephasing of the polarization components, but to the entanglement between the photon SAM and the OAM degrees of freedom. Suitable quantum erasing scheme could remove such entanglement so that the SISTOC component could be removed from the beam. We studied the SISTOC process in a TGG Faraday isolator, but the process is very general and it may occur in isotropic materials, as, for example, Nd:Glass rods under strong pumping conditions.


\begin{thebibliography}{30}
\bibitem{erasing} A quantum eraser based on C-NOT logic in the photon spinorbit space is under study and will be the object of a forthcoming work.
\bibitem{anisotropy}TGG is a ``weakly anisotropic'' cubic material, where the difference $p_{11}-p_{12}-2p_{44}$ among the photoelastic coefficients is very small (see \cite{Khazanov99} for notation).
\bibitem{reducedmatrix} In a maximally spinorbit entangled state, the reduced quantum density matrix (known in optics as the coherency matrix) of the SAM degree of freedom is proportional to the 2D unit matrix.
\bibitem{fidelity} The quantum fidelity is given by $\langle\psi|\rho|\psi\rangle$, where $\rho$ and $|\psi\rangle$ are the experimental density matrix and the expected theoretical state.
\end{thebibliography}
\end{document}